\documentclass[twocolumn,prd,noshowpacs,nofootinbib,amsmath,amssymb,superscriptaddress,preprintnumbers]{revtex4}

\usepackage{graphicx,epsfig,psfrag,bm,amssymb}
\usepackage{mathrsfs,amsmath, amsfonts,hepunits, color}

\usepackage{dcolumn}
\usepackage{bm}
\usepackage{mciteplus}
\usepackage{tikz}
\usepackage{slashed}
\usepackage{multirow}
\usepackage{slashed}

\definecolor{navy}{rgb}{0.2,0.2,0.7}
\usepackage[colorlinks]{hyperref}
\hypersetup{
     colorlinks  = true,
     citecolor  = navy,
 	linkcolor=red,
	urlcolor=navy
}

\newcommand{\be}{\begin{eqnarray}}
\newcommand{\ee}{\end{eqnarray}}
\def\lsim{\mathrel{\rlap{\lower4pt\hbox{\hskip 0.5 pt$\sim$}}
    \raise1pt\hbox{$<$}}}                
\def\gsim{\mathrel{\rlap{\lower4pt\hbox{\hskip1pt$\sim$}}
    \raise1pt\hbox{$>$}}}

\def\lsim{\mathrel{\rlap{\lower4pt\hbox{\hskip1pt$\sim$}}
    \raise1pt\hbox{$<$}}}
\def\gsim{\mathrel{\rlap{\lower4pt\hbox{\hskip1pt$\sim$}}
    \raise1pt\hbox{$>$}}}

\definecolor{Purple}{rgb}{0.8,0.2,0.5}

\begin{document}

\title{Central-edge asymmetry as a probe of Higgs-top coupling in $t\bar{t}h$ production at the LHC}
\author{Jinmian Li}
 \email{jmli@kias.re.kr}
\affiliation{School of Physics, KIAS, Seoul 02455, Korea}
\author{Zong-guo Si}
 \email{zgsi@sdu.edu.cn}
 \affiliation{School of Physics, Shandong University, Jinan, Shandong 250100, China}
\author{Lei Wu}
 \email{corresponding author:leiwu@itp.ac.cn}
 \affiliation{Department of Physics and Institute of Theoretical Physics, Nanjing Normal University, Nanjing, Jiangsu 210023, China}
   \affiliation{ARC Centre of Excellence for Particle Physics at the Terascale, School of Physics, The University of Sydney, NSW 2006, Australia}
\author{Jason Yue}
\email{jason.yue@ntnu.edu.tw}
 \affiliation{Department of Physics, National Taiwan Normal University, Taipei 116, Taiwan}
 \preprint{}

\begin{abstract}
The Higgs-top coupling plays a central role in the hierarchy problem and the vacuum stability of the Standard Model (SM). We propose a central-edge asymmetry ($A_{CE}$) to probe the CP violating Higgs-top coupling in dileptonic channel of $t\bar{t}h(\to b\bar{b})$ production at the LHC. We demonstrate that the CP-violating Higgs-top coupling can affect the central-edge asymmetry through distorting $\Delta y_{\ell^+\ell^-}$ distribution because of the contribution of new top charge asymmetric term. Since $\Delta y_{\ell^+\ell^-}$ distribution is frame-independent and has a good discrimination even in boosted regime, we use the jet substructure technique to enhance the observability of the dileptonic channel of $t\bar{t}h$ production. We find that (1) the significance of dileptonic channel of $t\bar{t}h$ production can reach $5\sigma$ for CP phase $\xi=0,\pi/4,\pi/2$ when the luminosity ${\cal L}=795,993,1276$ fb$^{-1}$ at 14 TeV LHC. (2) the central-edge asymmetry $A_{CE}$ show a good discrimination power of CP phase of $t\bar{t}h$ interaction, which are -40.26\%, -26.60\%, -9.47\% for $\xi=0$, $\pi/4$, $ \pi/2$ respectively and are hardly affected by the event selections. Besides, by performing the binned-$\chi^2$ analysis of $\Delta y_{\ell^+\ell^-}$ distribution, we find that the scalar and pseudo-scalar interactions can be distinguished at 95\% C.L. level at 14 TeV HL-LHC.
\end{abstract}

\maketitle

\section{Introduction}
After the discovery of the Higgs boson at the LHC \cite{higgs-atlas,higgs-cms}, precision study of its properties becomes one of the most important tasks in theory and experiment. So far, the measured Higgs gauge couplings are compatible with the SM predictions at 1-2$\sigma$ level. However, the Higgs fermion couplings remain obscure. Among them, the Higgs-top coupling is of particular interest.

In the SM, the top quark has the strongest coupling with the Higgs boson. As such, the Higgs-top coupling plays an special role in the hierarchy problem \cite{tHooft:1979rat} and the vacuum stability of the SM \cite{Sher:1988mj,Degrassi:2012ry}. Many models for physics beyond the SM related with these two problems predict a modified Higgs-top coupling. So, the precise measurements of Higgs-top coupling could give an insight on the pattern of fermion mass generation and the energy scale of new physics above the electroweak scale \cite{Bezrukov:2014ina}.

The most general Lagrangian of the $t\bar{t}h$ interaction can be parameterised as follows:
\begin{equation}
{\cal L} \supset -\frac{y_t}{\sqrt{2}}\overline{t}(\cos\xi+i\gamma_{5}\sin\xi)th,
\label{tth}
\end{equation}
where $y_t$ takes the value $y^{\rm SM}_t=\sqrt{2}m_t/v$ and $\xi=0$ in the SM~\cite{AguilarSaavedra:2009mx}, with $v=246$ GeV being the
vacuum expectation value of the Higgs field.

At the LHC, $t\bar{t}h$ production is the most promising direct way to probe the Higgs-top coupling \cite{Marciano:1991qq,Ellis:2013yxa,Kobakhidze:2014gqa,Khatibi:2014bsa,Yue:2014tya,Buckley:2015vsa,Boudjema:2015nda,Li:2015kaa,Gritsan:2016hjl,Dolan:2016qvg,Chang:2016mso,Cao:2016wib,Kobakhidze:2016mfx}. With the data set of 7 and 8 TeV runs of LHC, the signal strengths in the $t\bar{t}h$ production channel have been measured by both ATLAS \cite{Aad:2014lma,Aad:2015gra} and CMS \cite{Khachatryan:2014qaa} in various Higgs decay modes: $b\bar{b}$, $\tau^+\tau^-$ and $W^+W^-$. Given the large boosted cross section of $t\bar{t}h$ \cite{deFlorian:2016spz}, the LHC Run-2 would be able to pin down $t\bar{t}h$ production.

The favored channel for observing $t\bar{t}h$ production at the LHC exploits the dominant Higgs decay mode $h \to b\bar{b}$. In Ref.~\cite{Buckley:2013auc}, the observability of purely hadronic top decay channel of $t\bar{t}h(\to b\bar{b})$ has been demonstrated. In Ref.~\cite{Artoisenet:2013vfa}, the matrix element method was used to improve the sensitivity of $t\bar{t}h$ production at the LHC. On the other hand, due to the large multiplicity of jets, the fully hadronic top decay channel has the poor ability to unveil the nature of the Higgs-top coupling in Eq.~\ref{tth}. Therefore, it is necessary to explore the observability of $t\bar{t}h$ in other decay modes of top quarks. In Refs.~\cite{Bernreuther:2010ny,Bernreuther:2012sx,Bramante:2014gda,Bernreuther:2015yna,Aguilar-Saavedra:2014kpa,Godbole:2015bda}, various spin polarization/correlation observables in $t\bar{t}h$ production are proposed to probe the Higgs-top coupling. However, the discrimination power of those spin observables is easily reduced by the experimental kinematical cuts.

In this work, we investigate a central-edge asymmetry that arises from the rapidity difference of two leptons ($\Delta y_{\ell^+\ell^-}$) from the top quark decays in the dileptonic channel of $t\bar{t}h(\to b\bar{b})$ production at the LHC. We demonstrate the CP-violating Higgs-top coupling can affect the central-edge asymmetry through distorting $\Delta y_{\ell^+\ell^-}$ distribution because of the contribution of new top charge asymmetric term. Since $\Delta y_{\ell^+\ell^-}$ distribution is frame-independent and has a good discrimination even in boosted regime, we apply the jet substructure technique to enhance the observability of the dileptonic channel of $t\bar{t}h$ production without reducing the discrimination power of the central-edge asymmetry. \\

\section{Calculations and Results}
At the LHC, the dominant production of $t\bar{t}h$ is through the gluon fusion. The high order QCD and EW corrections to the $t\bar{t}h$ production have recently been studied  \cite{Dawson:1998py,Frederix:2011zi,Yu:2014cka,Demartin:2014fia,Frixione:2014qaa,Frixione:2015zaa,Maltoni:2016yxb,Broggio:2016lfj}. The presence of the CP violating Higgs-top interaction in Eq.~\ref{tth} will lead to the top quark charge asymmetry term in $t\bar{t}h$ production. To see this, we take the $s$-channel gluon fusion subprocess as example. Assuming incoming gluons momenta $q_1$ and $q_2$, outgoing top and antitop momenta $p_t$, $p_{\bar{t}}$, and Higgs momentum $p_h$, the amplitude is given by
\begin{eqnarray}
{\cal M}&&={\cal M}_1+{\cal M}_2 \nonumber\\
 && \propto \frac{\bar{u} (t)\Gamma_{t\bar{t}h}[(\slashed p_{t}+\slashed p_h)+m_t]\gamma_{\rho}v(\bar{t})}{(2q_1\cdot q_2)(m^2_h+2p_t\cdot p_h)}J^{\rho}_{\mu\nu}\epsilon^{\mu}_{1}\epsilon^{\nu}_{2},\nonumber\\
 &&-\frac{\bar{u}(t)\gamma_{\rho}[(\slashed p_{\bar{t}}+\slashed p_h)-m_t]\Gamma_{t\bar{t}h}v(\bar{t})}{(2q_1\cdot q_2)(m^2_h+2p_{\bar{t}}\cdot p_h)}J^{\rho}_{\mu\nu}\epsilon^{\mu}_{1}\epsilon^{\nu}_{2}
\end{eqnarray}
where $J^\rho_{\mu\nu}$ denotes the triple gluon interaction and $\Gamma_{t\bar{t}h}=(\cos\theta+i\gamma_{5}\sin\theta)$. Its contribution to the cross section of $t\bar{t}h$ production involves the factor $Tr(\slashed p_t \gamma_\sigma \slashed p_{\bar{t}} \gamma_\tau \gamma_5)$, which is asymmetric in the interchange of $t$ and $\bar{t}$ and will affect the kinematics of the decay products of the top/anti-top quark.

\begin{figure}[t!]
\includegraphics[width=8cm, height=5cm ]{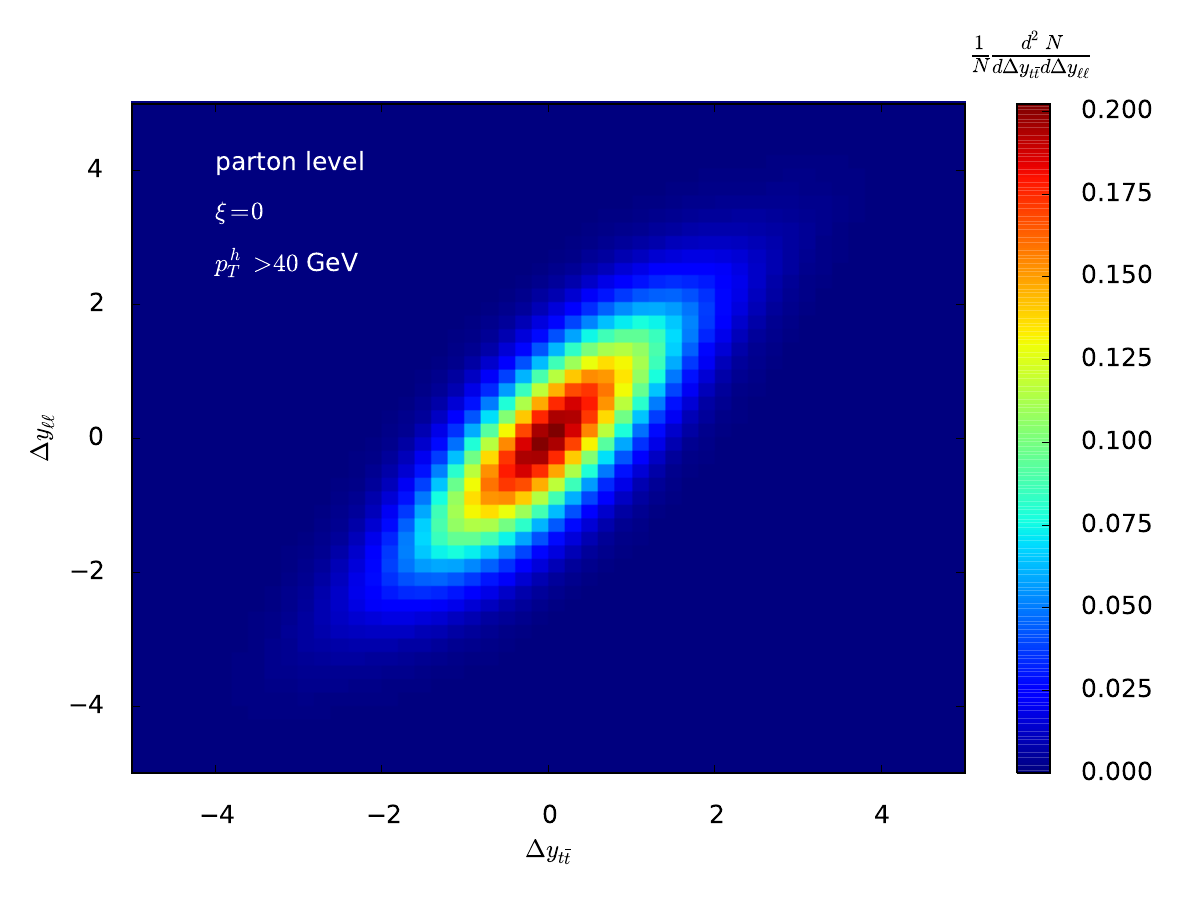}
\includegraphics[width=8cm, height=5cm ]{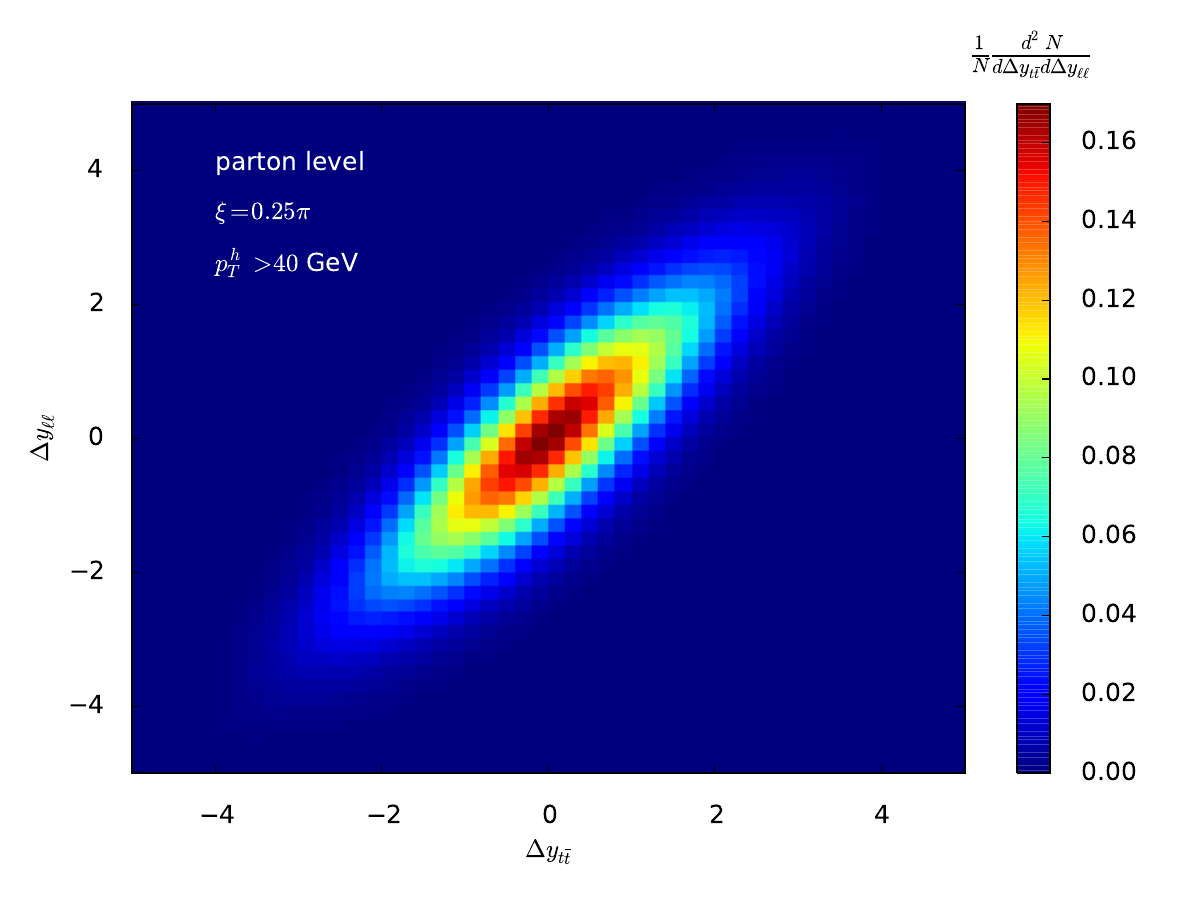}
\includegraphics[width=8cm, height=5cm ]{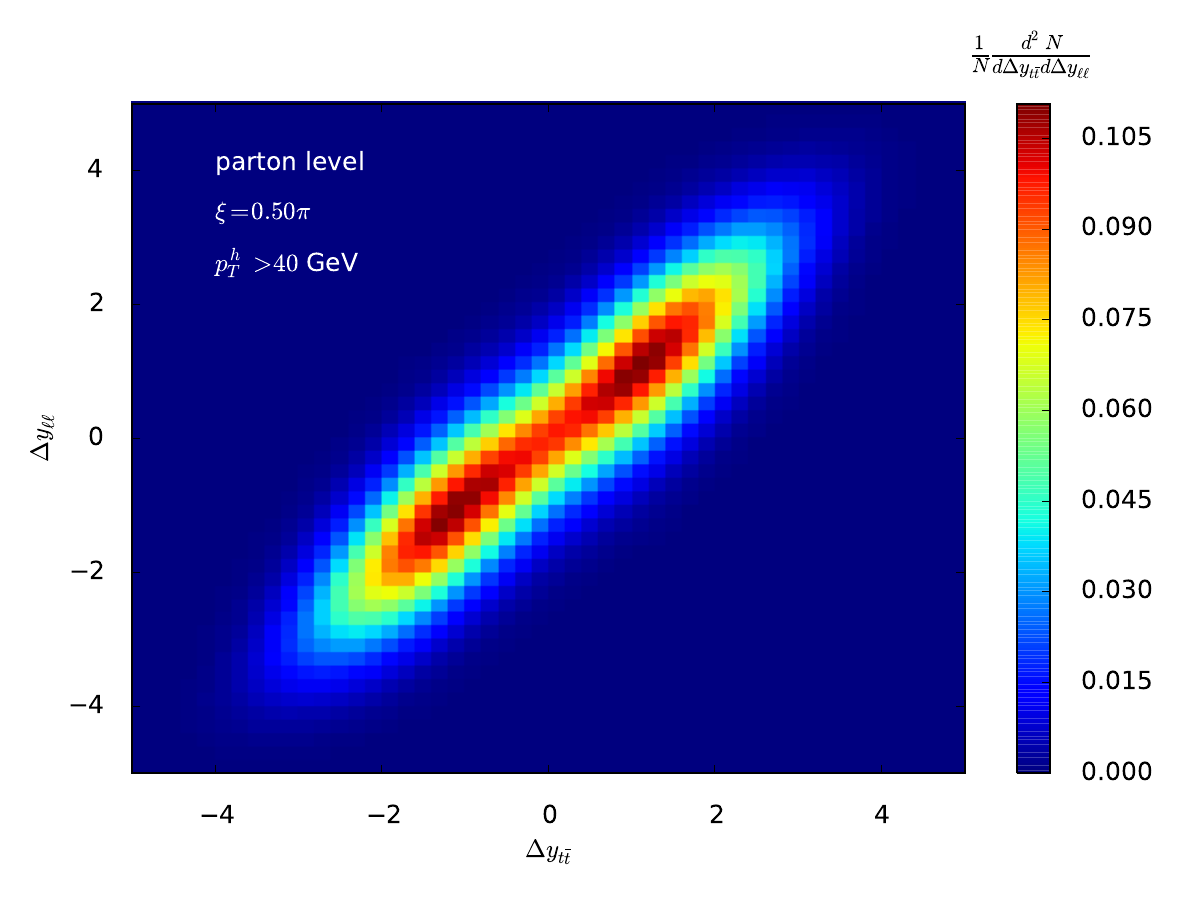}
\caption{Parton level correlation between $\Delta y_{\ell^+\ell^-}$ and $\Delta y_{t\bar{t}}$ in dileptonic $t\bar{t}h(\to b\bar{b})$ production for $\xi=0$ (upper), $\pi/4$ (middle), $\pi/2$ (lower) at 14 TeV LHC.}
\label{fig:correlation}
\end{figure}

In Fig.~\ref{fig:correlation}, we show the parton level correlations between $\Delta y_{\ell^+\ell^-}$ and $\Delta y_{t\bar{t}}$ in dileptonic $t\bar{t}h(\to b\bar{b})$ production for $\xi=0,\pi/4,\pi/2$ at 14 TeV LHC. We can see that $\Delta y_{\ell^+\ell^-}$ indeed has a strong correlation with $\Delta y_{t\bar{t}}$, which indicates that the dynamical reason for changing $\Delta y$ distribution comes from the above top quark charge asymmetric term rather than spin-correlation. For $\xi=\pi/4$ and $\pi/2$, the distributions of $\Delta y$ spreads towards the large values, as a comparison with $\xi=0$.

\begin{figure}[t]
\centering
\includegraphics[width=8cm, height=5cm ]{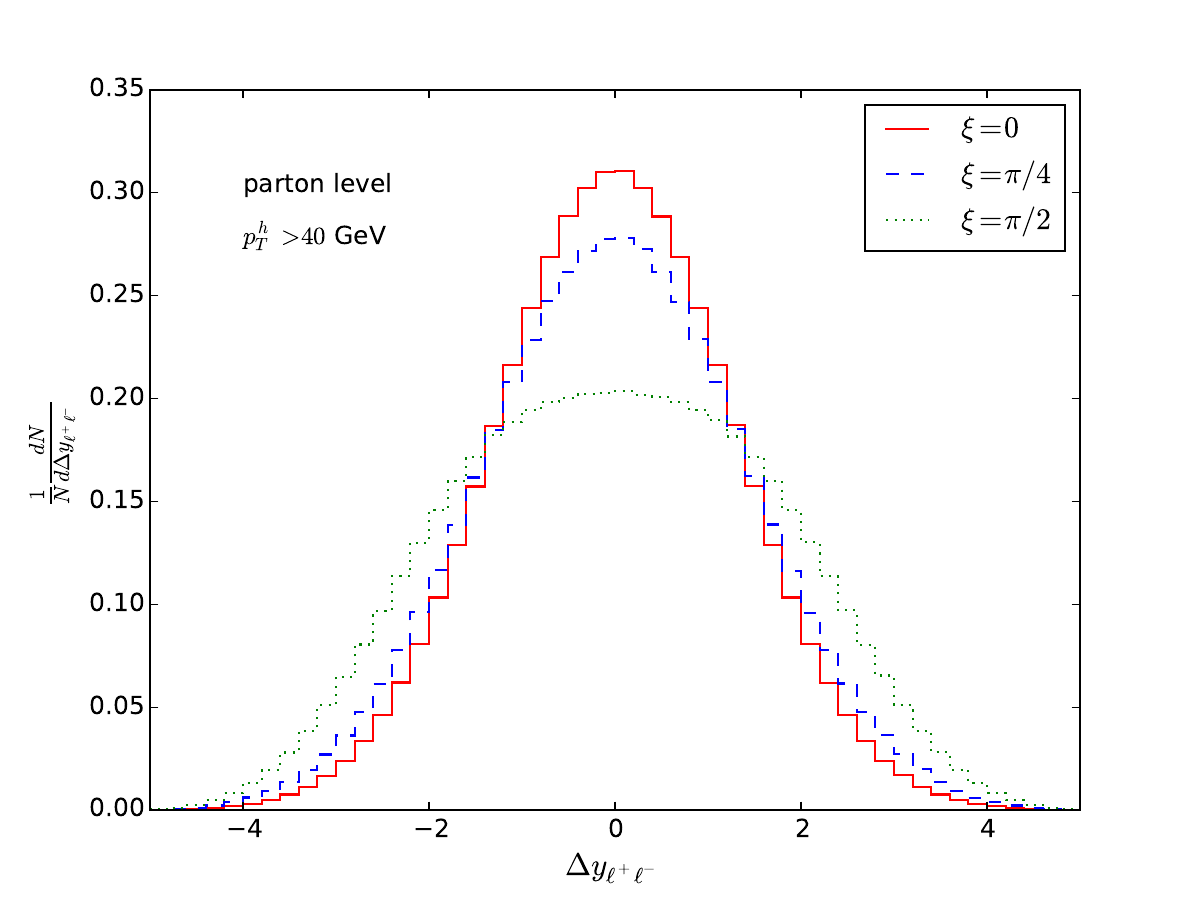}
\includegraphics[width=8cm, height=5cm ]{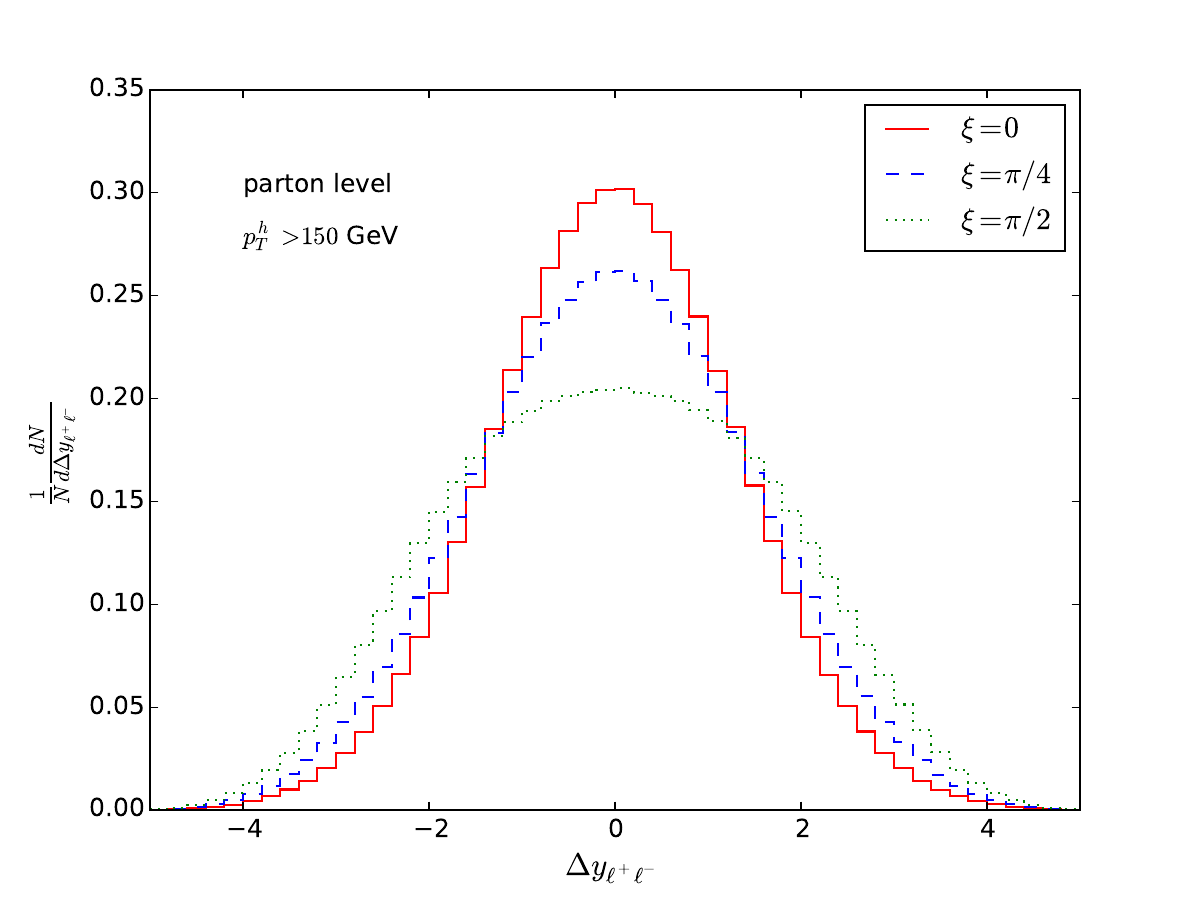}
\caption{Normalized parton-level $\Delta y_{\ell^+\ell^-}$ distribution in $t(\to b\ell^+ \nu_\ell)\bar{t}(\to \bar{b}\ell^-\nu_{\bar{\ell}})h$ production with $p^h_T>40$ GeV (upper panel) and $p^h_T> 150$ GeV (lower panel) at 14 TeV LHC.}
\label{fig:decay_tt_ll_1d}
\end{figure}
In Fig.~\ref{fig:decay_tt_ll_1d}, we present the parton-level distributions of $\Delta y_{\ell^+\ell^-}$ for $\xi=0, \pi/4, \pi/2$ in Eq.~\ref{tth} with $p^h_T>40$ and $150$ GeV at 14 TeV LHC. We can see that the the SM interaction ($\xi=0$) has more events than the mixed ($\xi=\pi/4$) interaction in the range of $|\Delta y_{\ell^+\ell^-}|<1.5$, followed by pseudo-scalar interaction ($\xi=\pi/2$). While the distribution is reverse in the range of $|\Delta y_{\ell^+\ell^-}|>1.5$. Such a behavior will give a small (large) asymmetry $A_{CE}$ for $\xi=\pi/2$ ($\xi=0$). Besides, it can seen that the difference among $\xi=0, \pi/4, \pi/2$ in $\Delta y_{\ell^+\ell^-}$ distribution is not sensitive to the increase of $p^h_T$. This indicates that the variable $\Delta y_{\ell^+\ell^-}$ has a good discriminating power for the different CP phases even in boosted phase space.

To quantitatively describe the difference in $\Delta y$ distributions for different CP phase, we define a central-edge asymmetry,
\begin{eqnarray}
A_{CE} \equiv \frac{\sigma_{|\Delta y_{\ell^+\ell^-}|>|\Delta y^0_{\ell^+\ell^-}|}-\sigma_{|\Delta y_{\ell^+\ell^-}|<|\Delta y^0_{\ell^+\ell^-}|}}{\sigma_{|\Delta y_{\ell^+\ell^-}|>|\Delta y^0_{\ell^+\ell^-}|}+\sigma_{|\Delta y_{\ell^+\ell^-}|<|\Delta y^0_{\ell^+\ell^-}|}},
\end{eqnarray}
where $\Delta y^0$ is the critical value of $\Delta y_{\ell^+\ell^-}$ and is determined from the crossing point of $\Delta y_{\ell^+\ell^-}$ distributions for the different CP phases. The prediction of $A_{CE}$ significantly different from the SM value of $t\bar{t}h$ production would strongly indicate the the non-standard CP violating Higgs-top interaction in Eq.~\ref{tth}.

\begin{table}[ht!]
\begin{center}
\begin{tabular}{|c| c|c| c|c|}\hline
  \multirow{2} {*}{~~$\xi$~~} &\multicolumn{2}{c|}{$A_{CE}(\ell^+\ell^-) (\%)$} \\\cline{2-3}
 &~~$p_T^h>40$ GeV ~~ &~~$p_T^h>150$ GeV~~\\ \hline
0        		& -52.00 	&-48.92\\
$\pi/4$  		&-41.13    &-35.58 \\
$\pi/2$ 	&-16.53 	&-16.73 \\ \hline
\end{tabular}
\caption{Parton-level values of $A_{CE}(\ell^+\ell^-)$ with $p_T^h>40,150$ GeV for $\xi=0,\pi/4,\pi/2$ at 14 TeV LHC.}\label{tab:prt_asym}
\end{center}
\end{table}
In Table~\ref{tab:prt_asym}, we numerically give the parton-level values of $A_{CE}(\ell^+\ell^-)$ for $\xi=0,\pi/4,\pi/2$ at 14 TeV LHC. For $p^h_T>40~(150)$ GeV, we can see that the value of $A_{CE}(\ell^+\ell^-)$ predicted by the SM is about -52\%(-49\%), while it becomes about -41\%(-36\%) and -17\%(-17\%) for the mixed and pseudo-interactions, respectively.

\begin{table*}[ht!]
\begin{center}
\begin{ruledtabular}
\begin{tabular}{c||c|c|c|c|c}
cut & $t\bar{t}h(\xi=0)$  & $t\bar{t}h(\xi=\pi/4)$ & $t\bar{t}h(\xi=\pi/2)$ & $t\bar{t}b\bar{b}$ & $t\bar{t}Z(\to b\bar{b})$ \\
\hline
2$\ell$, $p_{T}^{\ell} > 25 \ \text{GeV}$, $|\eta_{\ell}|<2.5$ & 13.31  &9.14 &5.31   & 2424.73 & 1.56\\\hline
$p_T^{\text{BDRS}}(b\bar{b}) > 150$ GeV & 2.02  & 1.47 &0.97    &19.24  &0.25 \\\hline
2 non-Higgs $b$'s & 0.28    & 0.21  & 0.15   &1.41 &0.04\\ \hline
$p_{T}^{b}(\text{non-}h) > 30 \ \text{GeV}$, $|\eta_{b}(\text{non-}h)|<2.5$   &0.22  &0.17 &0.13    &1.13 & 0.03 \\\hline
$\left| m_{b\overline{b}}^\text{BDRS} - 125\right|<10 $ GeV  &0.053   &0.048 &0.042   & 0.09  &0.0013
\end{tabular}
\end{ruledtabular}
\caption{Cut flow of the cross sections of the signal $t\bar{t}h$ for $\xi=0,\pi/4,\pi/2$ and backgrounds $t\bar{t}b\bar{b}$ and $t\bar{t}Z$ at 14 TeV LHC. The cross section is in unit fb.
\label{tab:cut_flow}}
\end{center}
\end{table*}

In the following, we study the observability of the dileptonic channel of $t\bar{t}h$ production with the sequent decay $h \to b\bar{b}$ and the charge asymmetry $A_{CE}(\ell^+\ell^-)$ for CP phases $\xi=0,\pi/4,\pi/2$ by including the detector effects at 14 TeV LHC. The dominant SM backgrounds are the $t\bar{t}b\bar{b}$ and $t\bar{t}Z(\to b\bar{b})$ productions. Since the signal and backgrounds have good discrimination in the high $p^h_T$ regime, we apply the jet substructure technique to reconstructing the Higgs boson.

We use \textsf{MadGraph5\_aMC@NLO}~\cite{Alwall:2014hca} to generate the parton-level signal and background events, in which the top quark and Higgs boson are further decayed with \textsf{Madspin}~\cite{Artoisenet:2012st}. The signal $t\bar{t}h$ and background $t\bar{t}Z$ is matched up to 1 jets by using MLM matching scheme \cite{mlm} with $xqcut=30$ GeV. We take $qcut$ to $max(xqcut+5, xqcut*1.2)$ \cite{matching} in our simulation. The CTEQ6M parton distribution functions (PDF) \cite{cteq} are chosen for our calculation. We set the renormalisation scale $\mu_R$ and factorisation scale $\mu_F$ to be $\mu_R=\mu_F=(m_h+2*m_t)/2$. We use \textsf{PYTHIA6}~\cite{Sjostrand:2006za} for implementing parton showering and hadronization. \textsf{Delphes3}~\cite{deFavereau:2013fsa} with input of default ATLAS detector card is used for simulating detector effects. In this simulation, we take the $b$-jet tagging efficiency as 70\% with the other light quark and gluon mis-tagging probability 1\%~\cite{Aad:2015ydr}.

Events which contain exactly two opposite sign leptons and at least four jets will be selected in our following analysis. These two leptons should have $p_T>15$ GeV, $|\eta|<2.5$ and be isolated. Particle-flow objects in \textsf{Delphes3} output other than isolated leptons are then used for jet clustering with \textsf{Fastjet}~\cite{Cacciari:2011ma}. We adopt the BDRS method for tagging Higgs jet substructure: (1) reconstructing the fat jets using C/A algorithm~\cite{Dokshitzer:1997in} with radius $R=1.5$ and $p^h_T >150$ GeV; (2) breaking each fat jet by undoing the clustering procedure. Higgs jet candidate is taken as the leading fat jet that has large mass drop $\mu<0.67$ and not too asymmetric mass splitting $y>0.09$ at certain step during the de-clustering; (3) filtering the Higgs neighbourhood by re-running the C/A algorithm with a finer angle $R_{filt} = \min(0.3,R_{j_1,j_2}/2)$ and taking the three hardest subjects; (4) applying $b$-tag on the two leading subjects. The Higgs jet candidate is required to have both subjects being $b$-tagged.  The pileup effects on the Higgs mass can be controlled by the BDRS filtering.
For event that contains the Higgs jet candidate, we proceed further to reconstruct narrow jets. The constituents of the Higgs jet candidate are removed from those particle-flow objects. The remnants are clustered with the anti-$k_T$ jet clustering algorithm~\cite{Cacciari:2008gp} with the cone radius of $R=0.4$ and are required to give at least two narrow jets, in which exactly two are $b$-tagged.

In Table~\ref{tab:cut_flow}, the cut-flow of cross sections of the signal and background events is presented for 14 TeV LHC. The cross sections of $t\bar{t}h$ are normalized to their NLO QCD values \cite{Demartin:2014fia}. After the cut $p^{BDRS}_T(b\bar{b})>150$ GeV, the $t\bar{t}b\bar{b}$ background is reduced by almost $\mathcal{O}(10^{-2})$, while the signals only by $\mathcal{O}(10^{-1})$. The Higgs mass window cut $|m^{\text BDRS}_{b\bar{b}}-125|<10$ GeV will further suppress $t\bar{t}b\bar{b}$ and $t\bar{t}Z$ backgrounds by one order. After all cuts, we find that the significance $S/\sqrt{B}$ of $\xi=0,\pi/4,\pi/2$ can reach $5\sigma$ when the luminosity ${\cal L}=795,993,1276$ fb$^{-1}$. The typical values of $S/B$ are about 30\%. The corresponding values of $A_{CE}(\ell^+\ell^-)$ are -40.26\%, -26.60\% and -9.47\%, which are mildly diminished by the event selections.

\begin{itemize}
  \item Cut $p^{BDRS}_T(b\bar{b})>150$ GeV, the $t\bar{t}b\bar{b}$ background is reduced by almost $\mathcal{O}(10^{-2})$, while the signals only by $\mathcal{O}(10^{-1})$.
  \item Cut $|m^{\text BDRS}_{b\bar{b}}-125|<10$ GeV will further suppress $t\bar{t}b\bar{b}$ and $t\bar{t}Z$ backgrounds by one order.
\end{itemize}


\begin{figure}[ht!]
\centering
\includegraphics[width=8cm, height=5cm ]{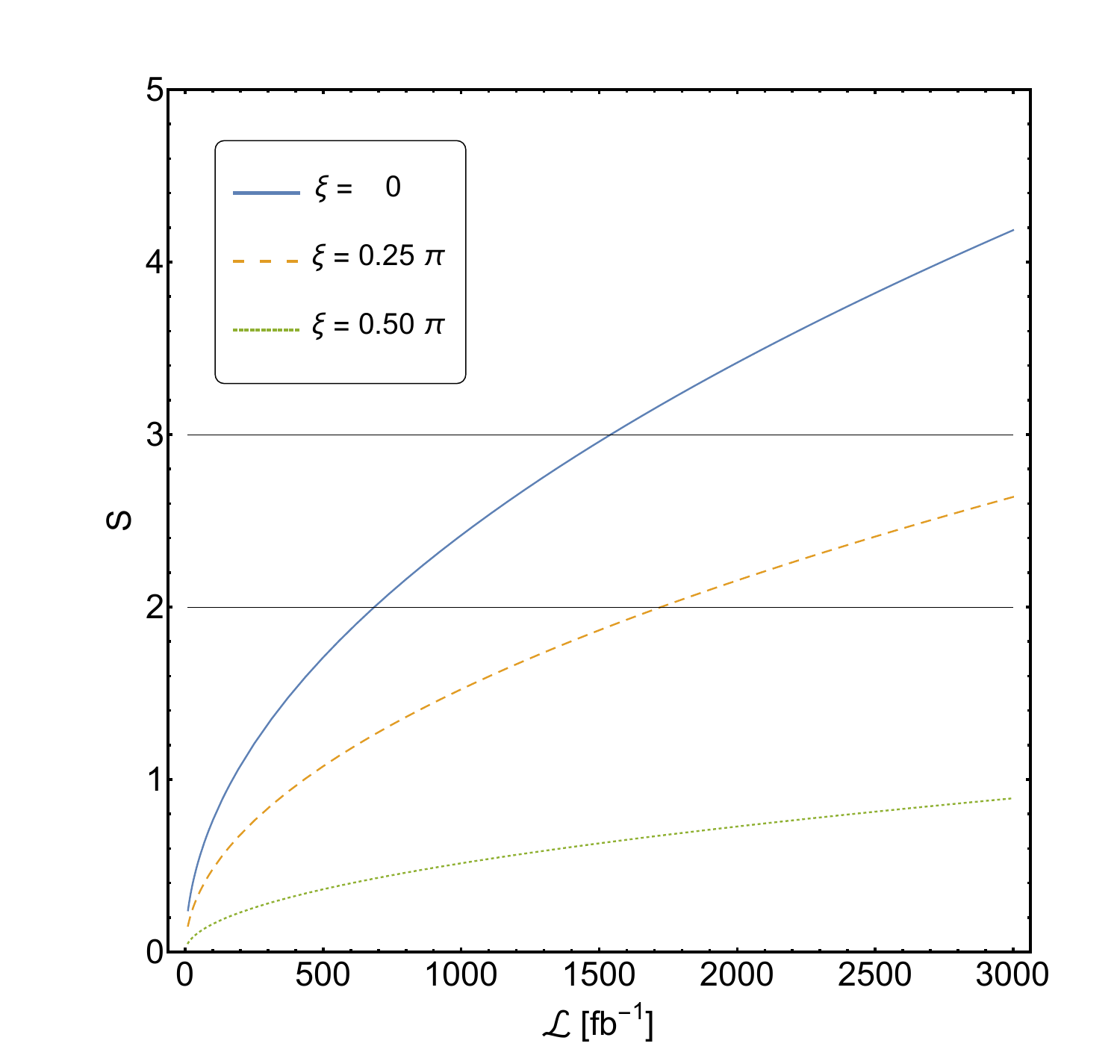}
\caption{The significance of $A_{CE}$ in dileptonic $t\bar{t}H(\to b\bar{b})$ production versus the integrated luminosity ${\cal L}$ for the CP phase $\xi=0,\pi/4,\pi/2$ at 14 TeV LHC.}
\label{fig:significance}
\end{figure}

A straightforward Gaussian estimate of the significance of $A_{CE}$ is given by
\begin{equation}
  S=\frac{A_{CE}}{\delta A_{CE}} \simeq \frac{|\Delta\sigma_{\Delta y_{\ell^+\ell^-}}|\mathcal{L}}{\sqrt{\sigma_{tot}\mathcal{L}}}.
\end{equation}
In Fig.~\ref{fig:significance}, we show the significance of $A_{CE}$ versus the luminosity $\cal L$ at 14 TeV LHC. We find that the SM prediction of $A_{CE}$ can be observed at $3\sigma$ level when ${\cal L}=1500$ fb$^{-1}$, while for the mixed and pseudo-scalar interactions, their significance is less than $3\sigma$ in the run of 14 TeV LHC.

\begin{figure}[ht!]
\centering
\includegraphics[width=8cm, height=5cm ]{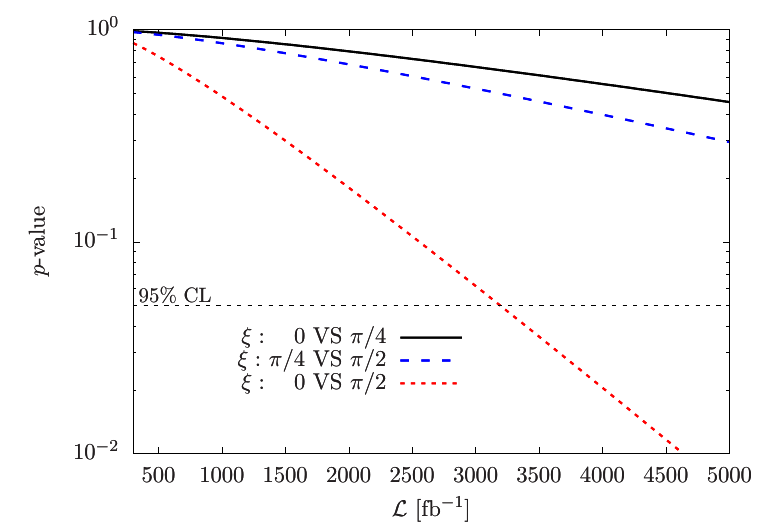}
\caption{The significance of $A_{CE}(\ell^+\ell^-)$ in dileptonic $t\bar{t}H(\to b\bar{b})$ production versus the integrated luminosity ${\cal L}$ for the CP phase $\xi=0,\pi/4,\pi/2$ at 14 TeV LHC.}
\label{fig:pvalue}
\end{figure}
Finally, we estimate the CP discrimination in Higgs-top couplings by calculating the binned-$\chi^2$ of the $\Delta y_{\ell^+\ell^-}$ histogram at reconstructed level. In Fig.~\ref{fig:pvalue}, we can see that the 14 TeV LHC will be able to distinguish $\xi=0$ and $\xi=\pi/2$ interactions at 95\% C.L. level if the luminosity ${\cal L} \simeq 3200$ fb$^{-1}$.

\section{Conclusions}
In this work, we investigate the CP violating Higgs-top couplings in dileptonic channel of $t\bar{t}h(\to b\bar{b})$ production at the LHC. We find that the CP violating interaction can distort the distribution of the rapidity difference of two leptons from the top decays because of the presence of the top quark charge asymmetric term. We also find that such an observable has a good discrimination power of the CP violating couplings in boosted regime. To numerically show the difference in $\Delta y_{\ell^+\ell^-}$ distributions, we define a central-edge asymmetry $A_{CE}$, which can reach -40.3\%, -26.6\% and -9.5\% for CP phase $\xi=0,\pi/4,\pi/2$, respectively. Besides, we simply perform the binned-$\chi^2$ analysis of $\Delta y_{\ell^+\ell^-}$ distribution and find that the scalar interaction and the pseudo-interaction can be distinguished at 95\% level at 14 TeV LHC with ${\cal L}\simeq 3200$ fb$^{-1}$ of integrated luminosity.

\acknowledgments
We thanks the valuable discussions with Mengchao Zhang, Archil Kobakhidze, J. M. Yang and C. R. Chen. This work is partly supported by the Australian Research Council (LW), by the National Natural Science Foundation of China (NNSFC) under grants Nos. 11325525 (ZS), by National Research Foundation of Korea (NRF) Research Grant NRF-2015R1A2A1A05001869 (JL).

\end{document}